\documentclass[pre,floats,superscriptaddress,showpacs,usenames]{revtex4}

\usepackage{cleveref}
\usepackage{array}
\usepackage{booktabs}
\usepackage{listings}
\usepackage{lmodern}
\usepackage{mathpazo}
\usepackage{microtype}
\usepackage{graphicx}

\usepackage{bm} 
\usepackage{amsmath}
\usepackage{amsbsy}
\usepackage{color}

\newcommand{\be}{\begin{equation}} 
\newcommand{\ee}{\end{equation}}

\newcommand{\bea}{\begin{eqnarray}}   
\newcommand{\eea}{\end{eqnarray}}

\newcommand{\ket}{\big>}
\newcommand{\bra}{\big<}

\newcommand{\ii}{{\bf i}}
\newcommand{\jj}{{\bf j}}

\newcommand{\bv}{{\bf v}}

\newcommand{\rr}{{\bf r}}
\newcommand{\NN}{{\bf \nabla}}
\newcommand{\FF}{{\bf F}}

\newcommand{\vv}{{\bf v}}

\newcommand{\fa}{f^{\alpha}}
\newcommand{\na}{n^{\alpha}}
\newcommand{\nb}{n^{\beta}}
\newcommand{\rhoa}{\rho^{\alpha}}

\newcommand{\cc}{{\bf c}}
\newcommand{\uu}{{\bf u}}
\newcommand{\uua}{{\bf u}^{\alpha}}

\newcommand{\uai}{u^{\alpha} _i}
\newcommand{\uaj}{u^{\alpha }_j}

\newcommand{\uub}{{\bf u}^{\beta}}
\newcommand{\ma} {m^{\alpha}}

\newcommand{\mb} {m^{\beta}}

\begin{document}
\date{\today}
\title{Lattice Boltzmann Method for mixtures at variable Schmidt number}

\author{Michele Monteferrante}
\address{Consiglio Nazionale delle Ricerche, Istituto di Chimica del Riconoscimento Molecolare (ICRM-CNR),
\\
Via Mario Bianco, 20131, Milan, Italy}
\author{Simone Melchionna}
\address{Istituto Processi Chimico-Fisici, Consiglio Nazionale delle Ricerche, Italy}

\author{Umberto Marini Bettolo Marconi\footnote[3]
{(umberto.marinibettolo@unicam.it)}
}
\address{ Scuola di Scienze e Tecnologie, 
Universit\`a di Camerino, Via Madonna delle Carceri, 62032,
Camerino, INFN Perugia, Italy}

\begin{abstract}
When simulating multicomponent mixtures via the Lattice Boltzmann Method, 
it is desirable to control the mutual diffusivity between species while maintaining the viscosity
of the solution fixed.  This goal is herein achieved by a modification of the multicomponent 
Bhatnagar-Gross-Krook (BGK) evolution equations by introducing two different timescales
for mass and momentum diffusion. Diffusivity is thus controlled
by an effective drag force acting between species.
Numerical simulations confirm the accuracy of the method for neutral binary and
charged ternary mixtures in bulk conditions.
The simulation of a charged mixture in a charged slit channel show that
the conductivity and electro-osmotic mobility exhibit a departure from the 
Helmholtz-Smoluchowski prediction at high diffusivity.
\end{abstract}
\maketitle

\section{Introduction}

In recent years, multicomponent transport is enjoying growing interest due to
burgeoning applications in biology, environmental science and energy 
management.  One of the most ubiquitous transport phenomena regards electrolytic solutions 
and involves multiple processes, including fluid flow, multi-species diffusion and
electrostatic interactions.
A better understanding of electrokinetic flows at nano and microscale level 
is paramount. At the nanoscale, for instance, it helps predicting 
mass and charge transport in biological ion channels.
At the microscale, it guides the design of devices for biomolecular diagnostics and 
energy transfer systems, like micro fuel cells and batteries. 
Electrokinetic flow is also important for non-mechanical actuating techniques, such
as for pumping, mixing and separating techniques \cite{schoch2008transport,kirby2010micro,berthier2010microfluidics}.

Computer simulation is a direct approach to investigate microfluidic and
nanofluidic systems and is exploited at best by those computational methods 
tailored to determine the concentration and current fields in generic confining geometries.
Among these, the Lattice Boltzmann Method (LBM) is a reference technique 
\cite{benzisuccivergassola,succi2001lattice}
and the reasons for its success are numerous:
i) LBM is based on kinetic theory and treats the elementary interactions 
between particles via either a lumped model or top-down approach 
or by resolving the collisional kinetics at microscopic, bottom-up level,
ii) the field dynamics is encoded by the single-particle distribution function and 
the governing equation is solved at second-order space/time accuracy by a compact numerical 
kernel, and
iii) being the support a simple cartesian mesh, the method is amenable to efficient and 
parallel implementation, with complex boundary conditions handled
in manageable and effective terms.
In general, the evolution of the distribution function can be driven by
several different interaction terms. Consequently, the hydrodynamic 
fields can reproduce a vast number of physical conditions. 
In electrokinetics the mixture is composed by a neutral solvent and at least 
two charged species, with the Poisson equation solved numerically 
for electrostatics under appropriate boundary conditions. 
When ions are dissolved in water, they display a diffusion coefficient $D\sim 10^{-9} m^2/s$
which is much smaller than the kinematic viscosity of the solution
(at ambient temperature pure water has $\nu=10^{-6} m^2/s$). 
In particular, charge transport displays an Ohmic contribution 
directly proportional to the diffusion coefficient, and a convective, 
electro-osmotic contribution inversely proportional to the kinematic viscosity. 
In order to account quantitatively for the two contributions, it is crucial 
that the numerical method reproduces such disparity in the transport coefficients. 
A related problem emerges when modeling the different timescales involved
in mass and momentum diffusion, where in a liquid the momentum field around a molecule 
diffuses much faster than the molecule itself. The Schmidt number, being the ratio between
kinematic viscosity and the diffusion coefficient ($Sc=\nu/D$), is as large as $10^2-10^3$ 
in aqueous solutions.

Perhaps the most popular implementation of the LBM is based on a reduction, 
proposed about sixty years ago by Bhatnagar-Gross-Krook (BGK) 
\cite{bhatnagar1954model},
of the Boltzmann transport equation. The BGK scheme replaces the 
collision kernel, involving integrals of products of the phase space 
distribution function, 
by a much simpler  form, describing the process of relaxation towards local equilibrium.
Such an approximate  treatment has certainly many advantages, 
both from the analytical point of view, because it allows to easily derive  hydrodynamic 
equations and closed expressions for the transport coefficients,
and from the computational point of view \cite{chen1998lattice}.
The BGK method, however, has an important limitation due to the fact that 
it is intrinsically built as a single-time relaxation process.  As a consequence,
the values of the mutual diffusion coefficient and kinematic viscosity
are exactly equal. 
Several authors have proposed alternatives to lift such restriction
by proposing, within the framework of a simplified kinetic model, 
tailored interactions among particles of different species
\cite{sirovich2004kinetic,hamel2004kinetic,hamel2004two,garzo1989kinetic}, 
but only few of them have addressed the practical problem of the dynamics of fluid mixtures,
apart from some notable exceptions 
\cite{sofonea2001bgk,luo2002lattice,guo2003discrete,asinari2008asymptotic}. 

Hamel proposed to control cross collisions by both an internal coupling force, 
proportional to the diffusion velocity, and an additional coupling term in the effective 
stress tensor \cite{hamel2004kinetic,hamel2004two}. 
Unfortunately, the mutual diffusivity and the mixture kinematic viscosity 
cannot be independently controlled within a single cross-collision relaxation time.
Subsequently, a multiple-relaxation-time (MRT) approach was proposed 
to independently control the mutual diffusivity and the mixture kinematic viscosity, 
the latter differing from the elementary mass averaged kinematic viscosity \cite{lallemand2000theory}.
However, if a BGK-like equation for each species is assumed, 
the equations does not recover the single component dynamics.
To overcome such drawback, Asinari proposed an alternative version within the MRT framework,
bridging the MRT scheme with the Hamel model \cite{asinari2008asymptotic}.
   
In the present paper, by using microscopically motivated physical arguments, we show 
how to take into account the occurrence of two different time scales associated 
with concentration and viscous momentum diffusion.
Such a separate control was previously achieved by considering the fully microscopic approach,
where the interactions between the particles were treated as hard-sphere collisions 
\cite{marconi2009kinetic,marconi2011dynamics}. However, the fully microscopic approach can be 
exceedingly demanding in computational terms and unneeded in practical situations.
Therefore, we propose a simpler version,  which can be viewed as a reduction of the 
full method, by using an effective drag force exerted between fluid components.
Inspired by kinetic density functional theory \cite{marconi2007phase}, 
we first introduce a microscopic description of the system, 
in which a free parameter $\omega_{drag}$  determines the diffusion coefficient. 
We then extend the description to ternary charged mixtures, generalizing
the self-consistent dynamical method from neutral binary mixtures 
\cite{marconi2011multicomponent} to electrolytic solutions. 
 
The paper is organized as  follows: 
in section \ref{Model} starting from a Boltzmann-like  transport approach
we introduce  a separation of the collision term
into a BGK-type  relaxation contribution plus a force term associated with the
drag forces between different fluid components which leads to a modified BGK 
multicomponent equation.
After deriving the balance equations for the momenta and the densities,
we show analytically that the diffusion and the viscosity coefficient display 
the required behavior. 
In section \ref{binary} we specialize the description to a neutral binary mixture, 
whereas in section \ref{ternary} we consider a ternary charged mixture and characterize 
their transport coefficients.
In section \ref{results} we corroborate the results by studying numerically 
the properties of the systems discussed above.
In section \ref{Conclusions} we make some conclusive remarks and considerations.

\section{Effective treatment of the kinetic equation }
\label{Model}

The exact time evolution of the one particle distribution function of the $\alpha$ species, $\fa(\rr,\vv,t)$, in a fluid mixture 
characterized by hard sphere interactions plus Coulomb forces can be formally written as
\bea
\frac{\partial}{\partial t}\fa(\rr,\vv,t) +\vv\cdot\NN \fa(\rr,\vv,t)
+\frac{\FF^{\alpha}(\rr)}{\ma}\cdot
\frac{\partial}{\partial \vv} \fa(\rr,\vv,t)
= \sum_{\beta}\Omega^{\alpha\beta}(\rr,\vv,t)
\label{equilproc}
\eea 
where $\ma$ are the particle masses, $\FF^{\alpha}$ an external force term and $\Omega^{\alpha\beta}(\rr,\vv,t)$ describes the effect of interaction between particles of type $\alpha$ and $\beta$ on the evolution of $\fa$.
When studying ternary charged mixtures, for instance, the index  $\alpha$ can take the values $\{-,0,+\}$ depending on the particle charge. 
 We assume that the interaction potential between two particles can be separated into a
short-range strongly repulsive and a long range  contribution (lr)
and correspondingly we make the approximation of splitting the collision term as:
\be
\Omega^{\alpha\beta}(\rr,\vv,t)=\tilde \Omega^{\alpha\beta}(\rr,\vv,t)+
\Omega^{\alpha\beta}_{lr}(\rr,\vv,t).
\label{opcoll}
\ee
The repulsive part  of $\tilde\Omega^{\alpha\beta}$ , assimilable to a chard-sphere potential, 
could be modeled using the Revised Enskog theory (RET) of Ernst and van Beijeren
\cite{van1973modified}
for hard-sphere mixtures
which  neglects velocity correlations for two particles about to collide, as in Boltzmann theory,
but
includes the configurational correlations resulting from the finite size of the particles via the pair correlation function at contact.
However, even the solution of RET equations represents a formidable numerical problem, so that a simplified treatment
is a valid alternative.
In fact, in the present paper we shall not treat in detail the hard sphere collision term, as we did in previous papers \cite{marconi2009kinetic,marconi2011dynamics}, and replace it by the much simpler BGK relaxation term.
  The  electrostatic contribution to the kinetic equation are 
treated in the mean field approximation, also known as Random Phase Approximation  \cite{hansen1990theory}, 
so that to rewrite  eq. \eqref{equilproc}
\bea
\frac{\partial}{\partial t}\fa(\rr,\vv,t) +\vv\cdot\NN \fa(\rr,\vv,t)
+\frac{\FF^{\alpha}(\rr)}{\ma}\cdot
\frac{\partial}{\partial \vv} \fa(\rr,\vv,t)
= \sum_{\beta}\bar \Omega^{\alpha\beta}(\rr,\vv,t)+\frac{e z^\alpha }{\ma}\NN \psi(\rr)\cdot
\frac{\partial}{\partial \vv} \fa(\rr,\vv,t)
\label{equilproc2}
\eea 
where $\psi(\rr)$ is the self consistent electric potential given by the solution of the Poisson equation 
 \be
 \nabla^2\tilde \psi(\rr)=-4\pi l_B[n^+(\rr)-n^-(\rr)]\;\;,\;\;\tilde\psi=e\psi/k_BT 
\label{Poisson}
 \ee
where $l_B=e^2/(4\pi k_BT\epsilon)$ is the Bjerrum length.
The boundary conditions for a surface charge density $\Sigma(\rr)$ are imposed as 
$[\nabla\tilde\psi(\rr)]_\perp=-{4\pi l_B\over e}\Sigma(\rr)$, where the symbol $\perp$ 
indicates the gradient component orthogonal to the surface. 
A practical treatment of the collision operator was suggested by Dufty et \emph{al} 
\cite{dufty1996practical,santos1998kinetic} who, starting from the revised Enskog theory 
(RET) for hard spheres systems \cite{van1973modified,van1973modified2}, 
proposed to separate the contributions to $\bar \Omega_{\alpha\beta}$
stemming from  the hydrodynamic modes from the non-hydrodynamic ones. 
Such a goal is achieved by projecting the collision term onto the
hydrodynamic subspace, spanned by the functions $\{1,\vv,\vv^{2}\}$, and onto the complementary kinetic subspace: 
\begin{equation}
\bar \Omega^{\alpha\beta}={\cal P}_{hydro}\bar \Omega^{\alpha\beta}+(I-{\cal P}_{hydro})\bar \Omega^{\alpha\beta}\, .
\label{splitting}
\end{equation}
In the case of constant temperature, the projection of the collision operator 
onto the hydrodynamic space is approximated by the following formula 
(see \cite{marconi2011multicomponent})
\bea
{\cal P}_{hydro}\bar \Omega^{\alpha\beta}=
{\phi^{\alpha}(\rr,t)\over \ma v_{T,\alpha}^2}
\Big[(\vv-\uu(\rr,t))\cdot \Phi^{\alpha}(\rr,t)\Big] 
\label{psia}
\eea
with $\phi^{\alpha}(\rr,\vv,t)=\na(\rr,t)[\frac{1}{2\pi v_{T,\alpha}^2}]^{3/2}\exp\Bigl(-\frac{(\vv-\uu(\rr,t))^2}{2 v_{T,\alpha}^2} \Big)$
being the local Maxwellian distribution and $v_{T,\alpha}=\sqrt{k_BT/m_{\alpha}}$  
the thermal velocity of the $\alpha$ species.

The terms $\Phi^{\alpha}(\rr,t)$ featuring in Eq. \eqref{psia} can be decomposed into different contributions stemming 
from different physical mechanism \cite{marconi2011non}:  a term proportional to the 
 gradient of the non ideal part of the chemical potential of species $\alpha$, a drag force, a viscous force and a force proportional to the
 gradient of the temperature.
For the present goal, we shall only consider the drag force acting on the  
$\alpha$ species originating by the presence of the $\beta$ species for slow varying densities. 
This contribution is proportional to
$
-\ma\gamma^{\alpha\beta}(\uua-\uub)
$
with
 $$\gamma^{\alpha\beta}= \frac{8}{3 \ma}\sigma_{\alpha\beta}^2 n^{\beta}\sqrt{\frac{ \ma\mb}{  \ma+\mb}\frac{2 k_B T}{\pi}} 
 g_{\alpha\beta} (\sigma_{\alpha\beta})  $$
 where $\sigma_\alpha$ is the molecular diameter of species $\alpha$,
$\sigma_{\alpha\beta}=(\sigma_\alpha+\sigma_\beta)/2$ and 
$g_{\alpha\beta} (\sigma_{\alpha\beta})$ is the pair correlation
 at contact  distance.
 In the case of molecules having approximately equal diameter $\sigma$ 
 and equal mass $m$, we can write
 \be
  \gamma^{\alpha\beta}=\frac{8}{3}\sigma^2 n \sqrt{ \frac{ k_B T}{m \pi}} g(\sigma) \frac{\nb}{n}
 \equiv\omega_{drag} \frac{\nb}{n}
 \ee  
where $\omega_{drag}$ has dimension of a frequency and will be used to tune the diffusion 
coefficient in the model.
 Hereafter, we shall assume  the following expression of the drag force $\FF^{\alpha,drag}$:
\bea
{\FF^{\alpha,drag}\over m}=-\omega_{drag}\sum_{\beta}{n_{\beta}(\rr,t)\over n(\rr,t)}[\uua(\rr,t)-\uub(\rr,t)] \, .
\eea
Concerning with the projection of $\bar\Omega_{\alpha\beta}$ onto the non-hydrodynamics 
sub-space, Dufty and coworkers, who dealt with the one component case only, 
approximated it by a  phenomenological  single relaxation-time BGK 
prescription, which  preserves the number of particles, the momentum and the kinetic energy 
and fulfills the physical symmetries and conservation laws of
the fluid \cite{dufty1996practical}.  Such a simple prescription,
when extended to multicomponent fluids,
has a serious drawback,  as stressed in \cite{marconi2011dynamics}, because 
it leads to the unphysical result that the diffusion coefficient and the kinematic viscosity 
have the same value.
In order to remedy such a situation we introduced the following  approximation for the decay of the non hydrodynamic modes:
\begin{equation}
(I-{\cal P}_{hydro})\bar \Omega^{\alpha\beta}\approx -\omega_{visc}[ \fa(\rr,\vv,t)- \phi^{\alpha}_{\perp}(\rr,\vv,t)] 
\label{dufty1}
\end{equation}
The "orthogonalized" Maxwellian distribution is defined as 
\bea
\phi^{\alpha}_{\perp}(\rr,\vv,t) &=& \phi^{\alpha}(\rr,\vv,t) \Bigl\{1+
{{[\uua(\rr,t)-\uu(\rr,t)]}\cdot {\cal H}^1\big(\vv-\uu(\rr,t)\big)\over v_{T,\alpha}^2}
\nonumber\\
&+&{1\over 2 v_{T,\alpha}^4 }[\uua(\rr,t)-\uu(\rr,t)][(\uua(\rr,t)-\uu(\rr,t)]:{\cal H}^2\big(\vv-\uu(\rr,t)\big)\Bigl\},
\label{prefactor}
\eea
where ${\cal H}^k$
are Hermite tensorial polynomials of order $k$, such that the kernel has a null projection 
onto the $\{1, \vv, \vv^2 \}$ subspace.
The modified collision operator \eqref{dufty1} contributes to determine the value
of the shear viscosity, which turns out to be a function of 
$\omega_{visc}$, a phenomenological collision frequency chosen to reproduce
the kinetic contribution to viscosity. 
The final result is the following set of  Enskog-like equations  
\begin{eqnarray}
\frac{\partial}{\partial t}\fa(\rr,\vv,t) +\vv\cdot\NN \fa(\rr,\vv,t) +
\frac{\FF^{\alpha}(\rr)}{m}\cdot
\frac{\partial}{\partial \vv} \fa(\rr,\vv,t)
&=& - \omega_{visc}[ \fa(\rr,\vv,t)- \phi^{\alpha}_{\perp}(\rr,\vv,t)]
\nonumber\\
&+&
\frac{{\bm\FF}^{\alpha,drag}(\rr,t) }{m v_{T}^2} \cdot(\vv-\uu(\rr,t)) 
 \phi^{\alpha}(\rr,\vv,t) 
\nonumber\\
&+&
\frac{e z^\alpha }{m}\NN \psi(\rr)\cdot
\frac{\partial}{\partial \vv} \fa(\rr,\vv,t) \, .
\label{evolution}
\end{eqnarray}
Notice that in the case of a one-component fluid there is no difference between $\phi^{\alpha}_{\perp}$ and $\phi^{\alpha}$,
since the velocities $\uua$ and $\uu$ coincide.
The above prescription fulfills the indifferentiability principle which states that
when all physical properties of the species are identical, the total distribution
$f=f^A+f^B$ must obey the single species transport equation.

The reason to use the modified distributions in Eq.~\eqref{prefactor} 
instead of Eq.~\eqref{psia}  
is to obtain the correct mutual diffusion and hydrodynamic properties starting 
from Eq.~\eqref{evolution}. 
From the knowledge of the $\fa$'s it is possible to determine not only all the hydrodynamic fields of interest, but
also the structure of the fluid at the molecular scale. 
The equations constituting the building blocks of the classical 
electrokinetic approach can be derived from the set of equations \eqref{evolution}.
In fact, the Poisson-Nernst-Planck (PNP) and Navier-Stokes (NS) equations 
\cite{bruus2008theoretical} are straightforwardly recovered 
by taking the appropriate velocity moments of the kinetic Enskog-Boltzmann equation. 
This derivation is performed in the next section in the limit of slowly varying fields. 

\section{Derivation of the equations of electrokinetics from the microscopic approach}
\label{macro}

Starting from the distributions functions $\fa(\rr,\vv,t)$, 
we define the partial number densities,
\be
\na(\rr,t)=\int d\vv \fa(\rr,\vv,t) \, ,
\ee
the mass densities $\rhoa(\rr,t)$,
\be
\rhoa(\rr,t)=m_{\alpha}\na(\rr,t),
\ee
the species velocities $\uua(\rr,t)$,
\be
\uua(\rr,t)= {1\over \na(\rr,t)} \int  \vv\fa(\rr,\vv,t)d\vv \, ,
\ee
the barycentric velocity,
\be
\uu(\rr,t)= \frac{\sum_{\alpha}\rhoa(\rr,t)\uua(\rr,t)} {{\rho(\rr,t)}},
\ee
{with $\rho(\rr,t)=\sum_\alpha \rhoa(\rr,t)$},  and the kinetic contribution of the $\alpha$ component to the pressure tensor
\be
\pi_{ij}^{\alpha}(\rr,t)=\ma\int d\vv (v_i-u_i)(v_j-u_j)\fa(\rr,\vv,t).
\label{pressurekin}
\ee
We now integrate Eq.~\eqref{evolution} w.r.t. the velocity and, assuming that the distribution functions $\fa(\rr,\vv,t)$ go to zero sufficiently fast, obtain the conservation law for the
particle number of each species
\be
\frac{\partial}{\partial t}\na(\rr,t) =-
\nabla\cdot \Bigl(\na(\rr,t)(\uua(\rr,t)- \uu(\rr,t)\Bigl)-\nabla\cdot \Bigl(\na(\rr,t) \uu(\rr,t)\Bigl)\equiv-\nabla \cdot {\bf J}^\alpha(\rr,t)  \, .
\label{continuityb}
\ee
In order to recover the PNP equation, we consider the momentum balance for the 
species $\alpha$ which is obtained after multiplying by $\vv$ and integrating w.r.t. $\vv$:
\begin{align}
&
\frac{\partial}{\partial t}[\na(\rr,t)\uaj(\rr,t)]+ 
\nabla_i \Bigl(\na(\rr,t) \uai(\rr,t) \uaj(\rr,t)
-  \na(\rr,t)(u^{\alpha}_i(\rr,t)-u_i(\rr,t))( u^{\alpha}_j(\rr,t)-u_j(\rr,t))\Bigl)=
\nonumber\\
&
-\nabla_i \frac{ \pi_{ij}^{\alpha}(\rr,t)}{\ma}+ \frac{F^{\alpha}_j(\rr)}
{\ma}\na(\rr,t)+
 \frac{ \FF^{\alpha,drag}_{j}(\rr,t)}{\ma}\na(\rr,t) -\frac{e z^\alpha}{m_{\alpha}}\na(\rr,t)\nabla_j \psi(\rr,t) \, ,
\label{momentcomponent}
\end{align}
One can verify that
the presence of the collision term \eqref{dufty1} does not affect explicitly the hydrodynamic balance equations
\eqref{continuityb} and
\eqref{momentcomponent}, thus allowing the decoupling of the concentration diffusion from the momentum
diffusion which is the goal of the present work.
It is convenient to introduce the chemical potential of the individual species, $\mu^{\alpha}$,
by the following equality:
\be
n^\alpha(\rr,t)\nabla_i \mu^{\alpha}(\rr,t) =\nabla_j \pi_{ij}^{\alpha}(\rr,t) \delta_{ij} \, .
\label{chemdef}
\ee

To derive the total momentum equation, we sum Eq.~\eqref{momentcomponent} over all components 
and obtain the following expression:
\begin{align}
\partial_{t}u_j(\rr,t)+ u_i(\rr,t)\nabla_i u_j(\rr,t)+\frac{1}{\rho}\nabla_i P_{ij}
+\frac{1}{\rho}\sum_{\alpha=\pm} e z^\alpha n^\alpha(\rr,t)\nabla_j \psi(\rr,t)
 -\frac{1}{\rho} \sum_{\alpha=0,\pm} n^\alpha(\rr,t) F^{\alpha }_j(\rr) =0 \, ,
\label{globalmomentumcont}
\end{align}
where we have introduced the total pressure tensor
$P_{ij}(\rr,t)= \sum_\alpha \pi_{ij}^{\alpha}(\rr,t)$. 
Notice that the sum of the drag forces vanishes in \eqref{globalmomentumcont} by the third principle of Dynamics.
By neglecting short range interactions, the pressure tensor is of kinetic nature only and
can be cast in the following form:
\be
P_{ij}(\rr,t)= P_{id}(\rr,t) \delta_{ij}-\eta \Bigl( \frac{\partial u_i}{\partial x_j}+ \frac{\partial u_j}{\partial x_i}-\frac{2}{3}
 \frac{\partial u_k}{\partial x_k} \delta_{ij}  \Bigl)
 \label{pressure}
 \ee
where the diagonal part is the ideal gas pressure of the mixture $P_{id}=k_B T\sum_\alpha \na$
and $\eta$ is the dynamical shear viscosity.
In the absence of electric and external fields Eq.~\eqref{globalmomentumcont}
is identical to the one describing a one component fluid. Thus, by applying 
a standard  Chapman-Enskog  analysis, not reported here for brevity 
\cite{chapman1970mathematical}, 
it is possible to show  that the viscosity
is related to the relaxation parameter $\omega_{visc}$,featuring in Eq.~\eqref{evolution}, as:
\bea
\eta=\frac{v_T^2}{\omega_{visc}} \sum_\alpha\rhoa(\rr),
\label{viscexp}
\eea
a result applying to  mixtures of arbitrary number of components.

\subsection{Diffusion and viscosity of a binary neutral mixture}
\label{binary}
Let us specialize the treatment to two neutral 
species  A and B, so that  $\psi=0$, and the drag force acting on the $A$ species due to the $B$ species is
\be
\frac{\FF^{A,drag}(\rr,t)}{\ma}=  -\omega_{drag}
{n^{B}(\rr,t)\over n(\rr,t)}
[\uu^A(\rr,t)-\uu^B(\rr,t)] 
\label{gammaab}
\ee
assuming equal masses, negligible variation of the densities  and 
negligible non linear terms, Eq.~\eqref{momentcomponent} reads 
\bea
\frac{\partial}{ \partial t} (\uu^A(\rr,t)-\uu^B(\rr,t)) +\omega_{drag}
 (\uu^A(\rr,t)-\uu^B(\rr,t)) = 
  -\frac{1}{m}\Bigl\{\nabla \mu^{A}(\rr,t)
-\nabla \mu^{B}(\rr,t) - \FF^{A}(\rr)+ \FF^{B}(\rr)  \Bigl\} \, .
 \label{wequation}
\eea
By neglecting the relative acceleration, we obtain 
\bea
\uu^A(\rr,t)-\uu^B(\rr,t) =- \frac{1}{m}\frac{1}{\omega_{drag}}\Bigl\{\nabla \mu^{A}(\rr,t)
-\nabla \mu^{B}(\rr,t)  - \FF^{A}(\rr)+ \FF^{B}(\rr)  \Bigl\} \, ,
\label{wapprox}
\eea
that, in absence of external forces ($\FF^{A}=\FF^{B}=0$), becomes 
\be
 \uu^A(\rr,t)-\uu^B(\rr,t)  =-\frac{1}{m \omega_{drag}} (\nabla \mu^{A}(\rr,t)
-\nabla \mu^{B}(\rr,t)) \, .
 \label{wapproxb}
\ee
Assuming also that the barycentric velocity vanishes, we rewrite Eq.~\eqref{continuityb} as 
\be
\frac{\partial}{\partial t}n^A(\rr,t) 
=\nabla\cdot 
\Bigl[\frac{n^A n^B}{n^A+n^B} \frac{1}{m\omega_{drag}}
\Bigl(  \nabla \mu^A(\rr,t) - \nabla \mu^B(\rr,t)\Bigl)
\Bigl] \, .
\label{continuity2}
\ee
If the total density $n^A+n^B=n_0$ is constant, so that $n^A=c n_0$ and $n^B=(1-c)n_0$  
and $\mu^{A(B)}$ are approximated by their ideal gas expressions, one finds
\be
\frac{\partial}{\partial t}c(\rr,t) \approx  { v_T^2\over   \omega_{drag}}\nabla^2 c(\rr,t)  
\label{continuity7}
\ee
so that the mutual diffusion coefficient is
\be
D ={v_T^2\over \omega_{drag}}.
\label{mutualdiffusion}
\ee

\subsection{The charged mixture: conductivity and mass flow}
\label{ternary}
Let us turn back to the ternary charged mixture.
Assuming that a steady current exists, we drop the non-linear terms in the velocities in 
the l.h.s. of Eq.~\eqref{momentcomponent}, in absence of other external forces 
$\FF^\pm=0$, the approximated force balance, obtained from Eq.~\eqref{momentcomponent}, reads 
\be
\nabla \mu^{\pm}(\rr,t) +e z^\pm \nabla  \psi(\rr,t)
\approx \FF^{\pm,drag} (\rr,t) .
\label{momentcomponent2}
\ee
The r.h.s. of  Eq.~\eqref{momentcomponent2} represents the  drag force exerted on the particles of type $\alpha=\pm$,
in reason of their different drift velocities. In dilute solutions  the
charged components are expected to experience a large friction arising only from the solvent 
while a negligible friction  from the oppositely charged species, so that we further approximate
\be
\frac{\FF^{\pm,drag} (\rr,t)}{m} = -\omega_{drag} \Big[{n^0(\rr,t) \over n(\rr,t)}(\uu^\pm(\rr,t)-\uu^0(\rr,t))+{n^\pm(\rr,t) \over n(\rr,t)}(\uu^\pm(\rr,t)-\uu^\mp(\rr,t))\Big]\simeq-\omega_{drag}[\uu^\pm(\rr,t)-\uu(\rr,t)]
\label{dragforce}
\ee
where we set $n\simeq n^0$,  that is, $n^\pm << n^0$. We also have that
\bea
&&J^\pm(\rr,t)=n^\pm (\rr,t) \uu^\pm (\rr,t) =
n^\pm (\rr,t) (\uu^\pm (\rr,t) -\uu (\rr,t) )+n^\pm\uu(\rr,t) 
\approx  
n^\pm(\rr,t) \uu (\rr,t) -{1\over \ m \omega_{drag}}n^\pm (\rr,t) \FF^{\pm,drag} (\rr,t)  \, . \nonumber\\
\eea
Eliminating $\FF^{\pm,drag}$ in Eq.~\eqref{momentcomponent2}, the ionic current 
in the stationary state is:
 \be
{\bf J}^\pm(\rr,t)=  -{1 \over m \omega_{drag}} n^\pm(\rr,t) \nabla \mu^{\pm}(\rr,t) 
-\frac{1}{ m \omega_{drag}}   e z^\pm n^\pm(\rr,t)\nabla \psi(\rr,t)  + n^\pm(\rr,t) \uu(\rr,t) \, .
\label{microcurrent}
\ee
Eq.~\eqref{microcurrent} is the phenomenological Planck-Nernst current, which is
the sum of diffusive, migration and convective terms. The total electric charge density current is
\be
{\bf J_e}=
 -\sum_\pm \frac{e z^\pm}{m^\pm \omega_{drag}}  n^\pm(\rr,t)\nabla \mu^\pm(\rr,t) +  \sigma_{el} {\bf E}  +e\sum_{\pm} z^\pm
 n^\pm(\rr,t) \uu(\rr,t) \, ,
\ee
where the zero frequency electric conductivity $\sigma_{el}$ is given by the Drude-Lorentz 
like formula
\be
\sigma_{el}=\frac{e^2D}{v_T^2}  \Bigl( (z^+)^2 n^+ + (z^-)^2  n^- \Bigl) \, .
\label{Drude}
\ee
This expression shows that the conductivity is modulated by collisions 
with the solvent and decreases as the solvent becomes denser
($\omega_{drag}$ is an increasing function of $n^0$) while increases with the number of 
charge carriers.
For completeness we write  the macroscopic equation describing the electro-osmotic flow 
in the $x$-direction in a slit channel whose walls are normal to the $z$-axis, 
the so-called Stokes-Smoluchowski equation
\bea
&&\nabla_x P
+\epsilon E_x \frac{\partial^2 \psi(\rr,t)}{\partial z^2}     -\eta  \frac{\partial^2 u_x(\rr,t)}{\partial z^2}=0
\nonumber\\
\label{globalmomentumcontc}
\eea
and the Gouy-Chapman equation for the electric potential, which is based on the hypothesis 
that the charge density profile is in equilibrium in the direction orthogonal to the fluid flow,
\be
\frac{d^2}{d z^2}\psi(z)=\frac{e n}{\epsilon} \sinh{\frac{e  \psi(z)}{k_B T}}.
\label{gceq}
\ee
In the next section we shall compare the analytic predictions of 
eqs. \eqref{globalmomentumcontc} and \eqref{gceq} with the numerical solutions 
of the kinetic model.

\section{Results}
\label{results}

We performed simulations on binary neutral and ternary charged systems based on the
presented method.
In the discrete LBM representation, the standard procedure 
shows that the kinetic equation encoded by eq. \eqref{evolution} reads \cite{shan2006kinetic}
\bea
f_p^{\alpha}(\rr+\cc_p, t+1)=  f_p^{\alpha}(\rr,t)-
\omega_{visc} [f_p^{\alpha}(\rr,t)-\phi^{\alpha}_{p,\perp}(\rr,t)] 
+ {{\bf g}_p^{\alpha}(\rr,t)}
\label{latboeq}
\eea
where a unit time step is used and the vectors $\{\cc_p\}$ are a set of discrete velocities
used to sample the distribution in special points of velocity space.
For the latter set, we employ here the so-called D3Q19 velocity discretization 
\cite{succi2001lattice}.

The second term on the right side of Eq~\eqref{latboeq} describes the relaxation towards 
the modified local equilibrium $\phi^{\alpha}_{p,\perp}$,  
while ${{\bf g}_p^{\alpha}(\rr,t)}$ represents the contribution due to the forces acting on species 
$\alpha$. As shown in the Appendix,
both terms arise from the Hermite expansion of the collisional kernel 
and are taken up to second order truncations 
of the Hermite series representation.
The Chapman-Enskog analysis shows that the mutual diffusion coefficient and 
the kinematic viscosity are related to the input relaxation frequencies as
$D=v_T^2(1/\omega_{drag}-1/2 )$ and  
$\nu=v_T^2(1/\omega_{visc}-1/2)$, respectively. 
In the simulations, we always set $\nu=1$.  
 
\begin{figure}[htbp]
\begin{center}
\includegraphics[scale=0.8]{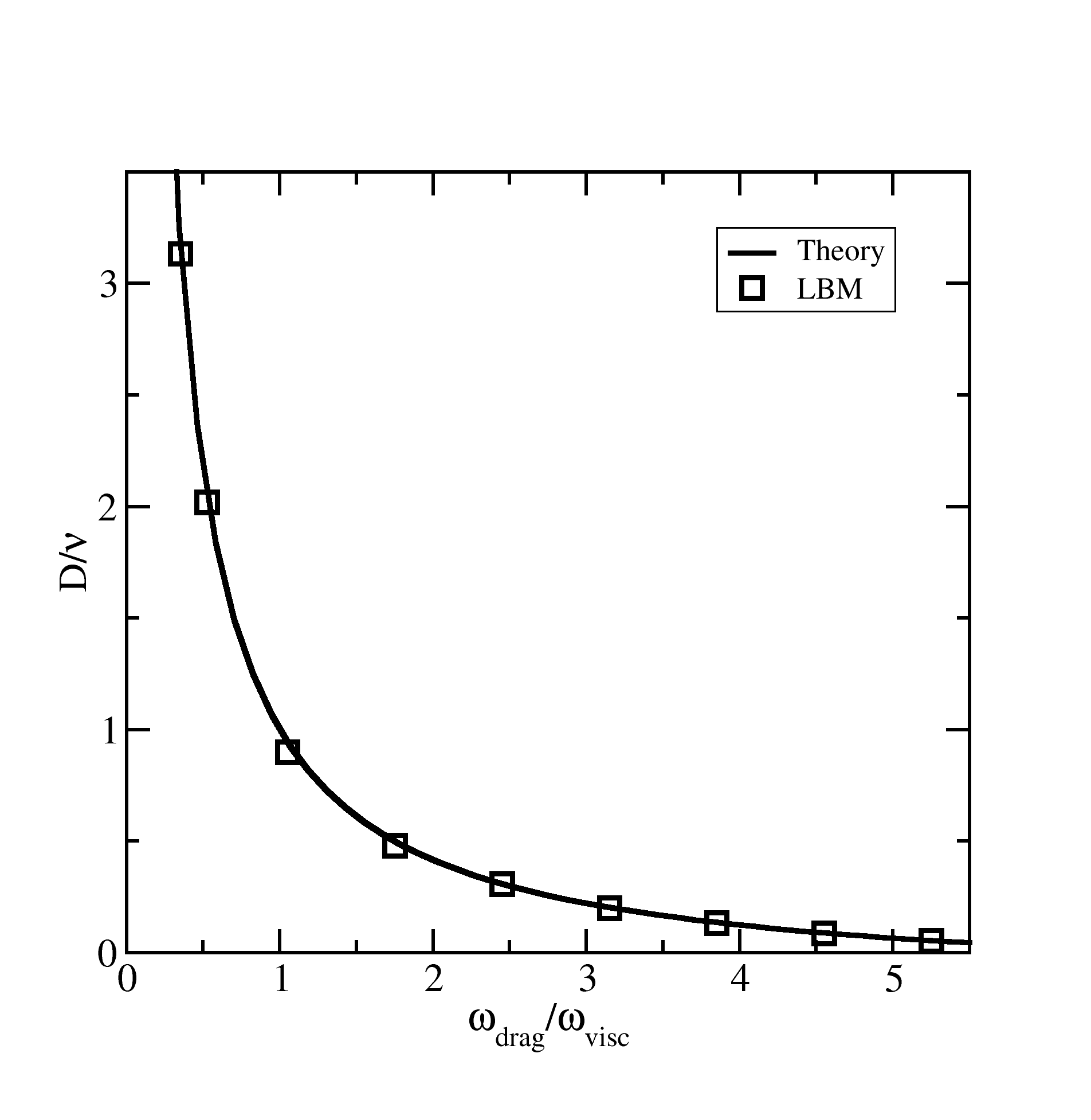}
\caption{Mutual diffusion coefficient for a neutral binary mixture in bulk conditions
vs $\omega_{drag}/\omega_{visc}$. 
Solid line: theoretical curve,
$D/\nu=(1/ \omega_{drag}-1/2)/(1/ \omega_{visc}-1/2)$,
symbols: LBM data.
}
\label{diffvsomega}
\end{center}
\end{figure}

We preliminarily tested the analytical predictions
of Eq.~\eqref{mutualdiffusion} for the mutual diffusion coefficient and 
of Eq.~\eqref{Drude} for the electric conductivity in bulk conditions. 
In these tests we used $L_x\times L_y\times L_z=50\times50\times50$  mesh points 
and applied periodic boundary conditions.  
To calculate the mutual diffusion coefficient of the binary mixture
we prepared the uniform system by adding a small sinusoidal concentration profile,
and monitored its exponential decay, having a wavelength dependent characteristic time  
$1/\tau(q_z)=D q_z^2$.
The diffusion coefficient was obtained a function of $\omega_{drag}$ 
and reported in Fig.~\ref{diffvsomega}.
The agreement with the theoretical values is excellent and the method allows 
to appreciably decrease the diffusion coefficient. 
To compute the kinematic viscosity we induced a sinusoidal shear modulation 
of the barycentric velocity, and monitored its decay. We found that upon changing the mutual 
diffusion coefficient, the kinematic viscosity remains unaltered, as expected. 

\begin{figure}[htbp]
\begin{center}
\includegraphics[scale=.8]{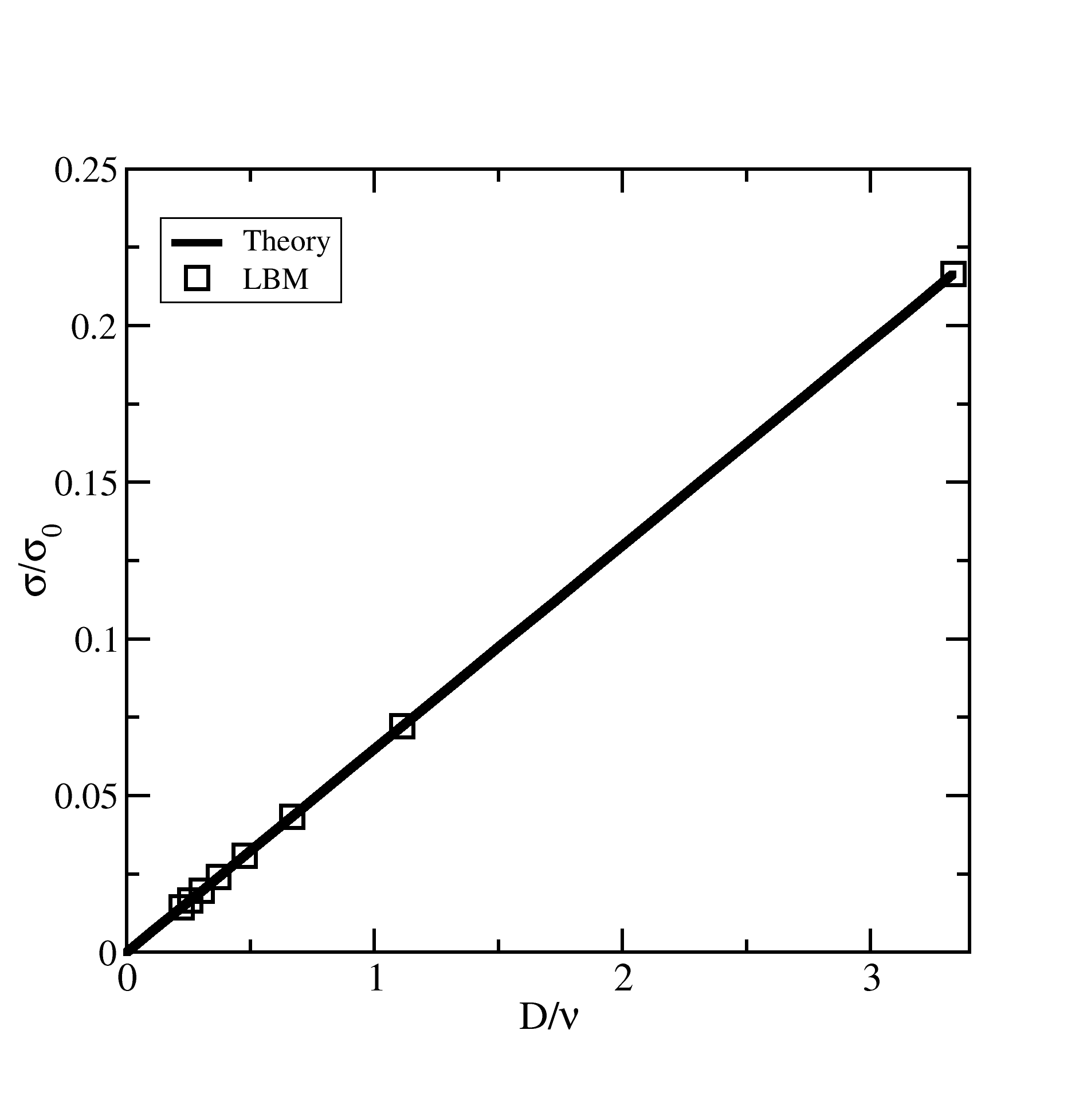}
\caption{Electrical conductance of a ternary charged mixture in bulk conditions
vs the mutual diffusion coefficient. 
The conductance is divided by the reference value 
$\sigma_0 = e (n^+ + n^-) v_T L_y L_z / E_x L_x$.
Solid line: Drude-Lorentz formula, Eq.~\eqref{Drude},
symbols: LBM data.
}
\label{sigvsdiff}
\end{center}
\end{figure}

We next turn our attention to the charged ternary mixture. 
We first conducted a test to compute the electric conductivity in the bulk system subject 
to an uniform electric field. 
The density of the electrolytes is taken to be $n^{\pm}/n^0=10^{-2}$.
We measured the total electric current at varying diffusivity and at constant electric field.
The data reported in Fig. ~\ref{sigvsdiff} show that the current is proportional 
to the diffusion coefficient $D$, as predicted by the Drude-Lorentz formula Eq.~\eqref{Drude}.

We next analyze the transport behavior of the electrolytic solution moving in a  slit 
channel having charged walls, under the action of a uniform electric field 
parallel to the walls. 
The flow is directed along the $x$ axis, 
the charged plates are aligned along the $xy$ plane, with surface area $L_x \times L_y$,
and the charged plates are separated along the z axis by $L_z$.
The total mesh size is $L_x \times L_y \times L_z = 40 \times 10 \times 50$. 
The surface charge of the walls is $\Sigma L_x L_y/e=-0.04$. 

From the macroscopic arguments used to derive the Hemholtz-Smoluchowski 
theory \cite{masliyah2006electrokinetic} one expects that the mass current, $I_{m}$ 
does not depend on the diffusion constant.
The bulk densities of the charged species $n^\pm_b$ are fixed by imposing the Debye length, 
since $\lambda_D = 1/\sqrt{4\pi l_B (n^+_b+n^-_b)}$,
and we considered the three values $\lambda_D/L_z=0.4$, $0.2$ and $0.1$.

\begin{figure}[htbp]
\begin{center}
\includegraphics[scale=.8]{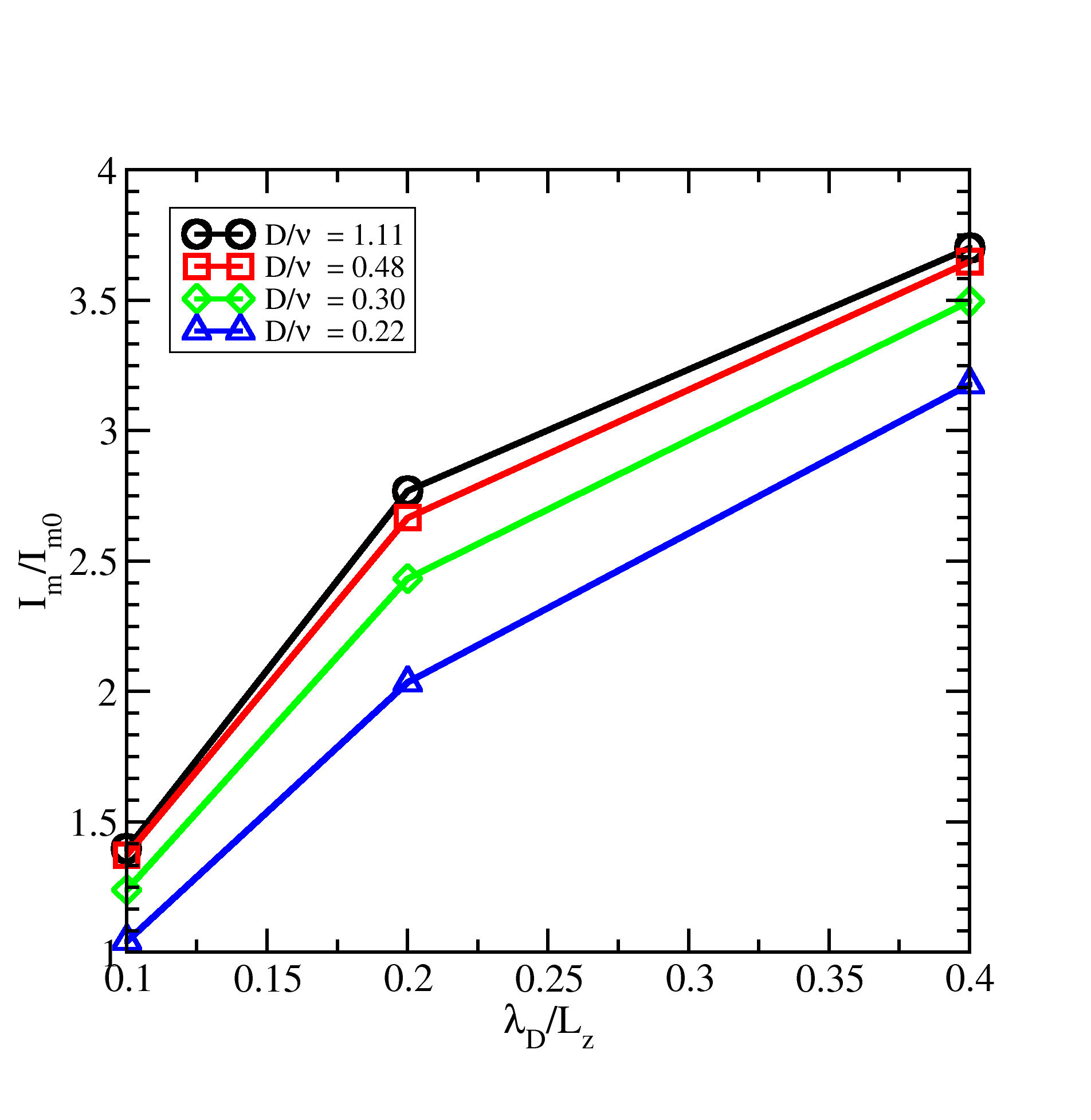}
\caption{Mass current vs $\lambda_D$  for the charged slit channel
obtained with LBM and computed via Eq.~\eqref{currmass}.
The mass current is divided by the reference value
$I_{m0} = E_x \Sigma L_x L_y / \nu $.
The curves are a guide to the eye.
 }
\label{massflow}
\end{center}
\end{figure}

The total mass flow rate in the stationary state is defined:
\be
I_m=m \int_{0}^{{L_z}} [n^+(\rr)u^+_x(\rr)+n^-(\rr)u^-_x(\rr)+n^0(\rr)u^0_x(\rr)] dz=m \int_{0}^{L_z}  n(\rr)\uu_x(\rr) dz
\label{currmass}
\ee
By neglecting small variations of the total density with respect to the bulk value, 
the theoretical mass flow rate is given by \cite{masliyah2006electrokinetic}
\be
I^{th}_m={E_x \Sigma\over \nu}\frac{1}{k_D^2} \Big[ 1- \frac{k_D L_z/2}{\tanh(k_D L_z/2)} \Big]
\label{theomassflow}
\ee
where $k_D=1/\lambda_D$ is the the inverse of the Debye length.
The quantity  $I_m^{th}$ is  an increasing function of $\lambda_D$, 
explaining the results shown in Fig.~\ref{massflow}. 
However, in the same plot we also notice a dependence of the mass flow rate on
$\omega_{drag}$. 
This behavior is not accounted for by the Helmholtz-Smoluchowski theory, which
in fact shows no dependence on the inter-species diffusivity.
However, it should be borne in mind that one of the key assumptions of the theory
is that the barycentric velocity locally equals the velocity of the neutral species, 
a condition that can be violated in the general case.

\begin{figure}[htbp]
\begin{center}
\includegraphics[scale=.8]{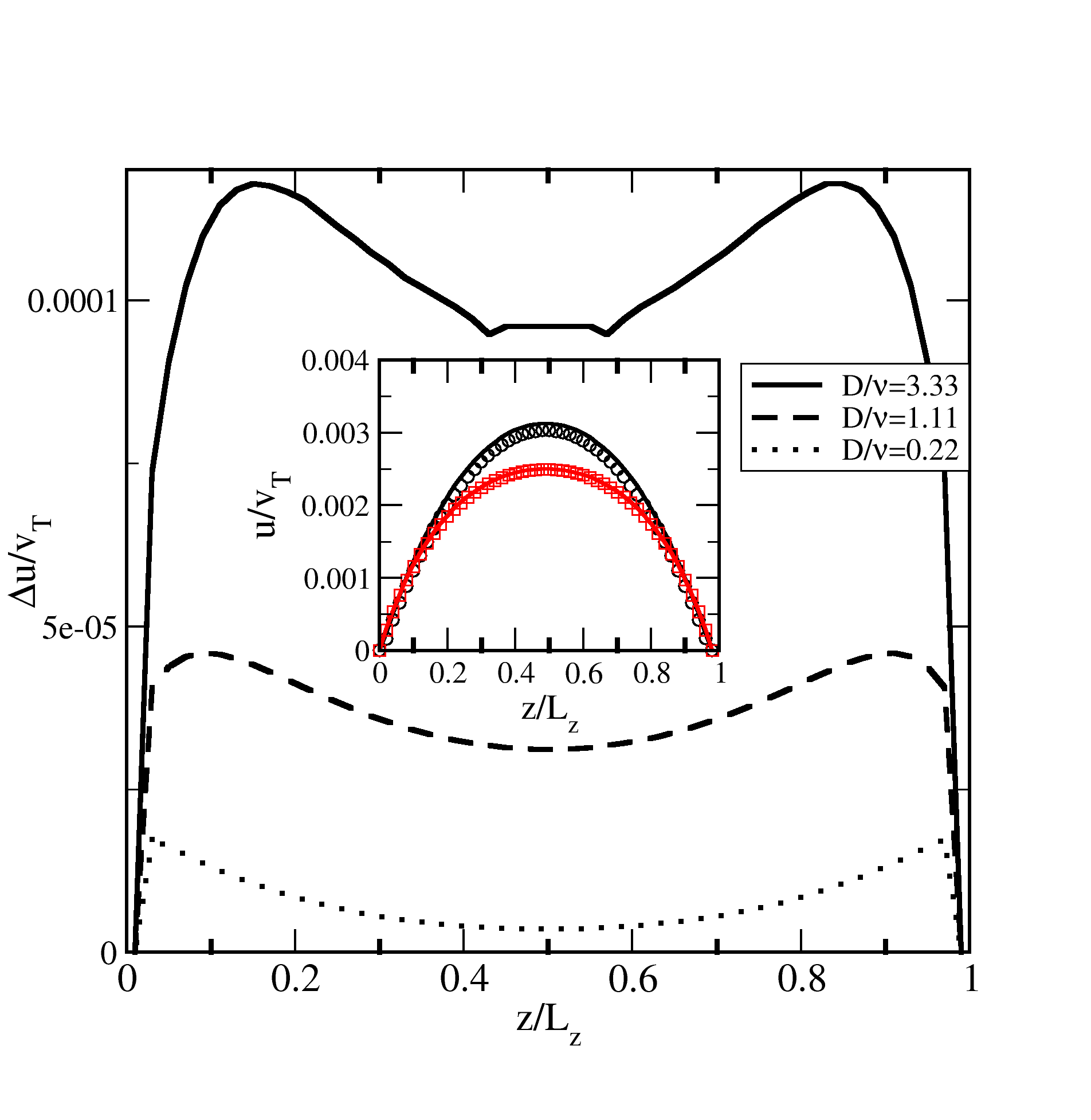}
\caption{Difference between the barycentric velocity and that of the neutral species 
for three diffusion coefficients as obtained via the LBM.
Inset: velocity of the neutral species (solid line) and the barycentric one (symbols) 
corresponding to the two limiting cases 
$\omega_{drag}/\omega_{visc}=0.35$ (circles) and $5.25$ (squares) and for $\lambda_D/L_z=0.1$.
}
\label{diffvel}
\end{center}
\end{figure}

To verify such assumption, we report in Fig.~\ref{diffvel} the differential velocity
profiles of the three species and find that, when $\omega_{drag}$ is large, 
the velocity of the neutral species is indistinguishable 
from the barycentric one. 
Vice versa, when the drag force is small the two velocities substantially 
depart from each other.

The electro-osmotic contribution to the charge flow rate displays a similar dependence 
on $\omega_{drag}$. To this purpose, we first consider the total electric current 
\be
I=e\int_{0}^{L_z} [n^+(\rr)u^+_x(\rr)-n^-(\rr)u^-_x(\rr)] dz,
\ee
and decompose it as the sum of two contributions stemming from 
Ohmic conduction and charge convection, $I=I_{Ohm}+I_{eo}$, with the latter defined as
\be
I_{eo}=e\int_{0}^{L_z}  [n^+(\rr)-n^-(\rr)]u_x(\rr) dz
\label{curreo}
\ee
and again $u_x$ is the barycentric velocity along the flow direction. 
Similarly, the electro-osmotic contribution to conductivity is obtained by dividing 
the electro-osmotic current by the applied tension.

In the linear (Debye-Huckel) approximation, the convective contribution can be computed
analytically and reads
\be
I^{th}_{eo}=-\frac{E_x \Sigma^2}{\eta  k_D \tanh^2(k_D L_z/2) }   \,\Big[ \frac{ \sinh(k_D L_z)+k_D L_z}{4\cosh^2(k_D L_z/2)}- \frac{ \sinh(k_D L_z/2)}{\cosh(k_D L_z/2)}\Big] \, .
\label{theo}
\ee

\begin{figure}[htbp]
\begin{center}
\includegraphics[scale=.8]{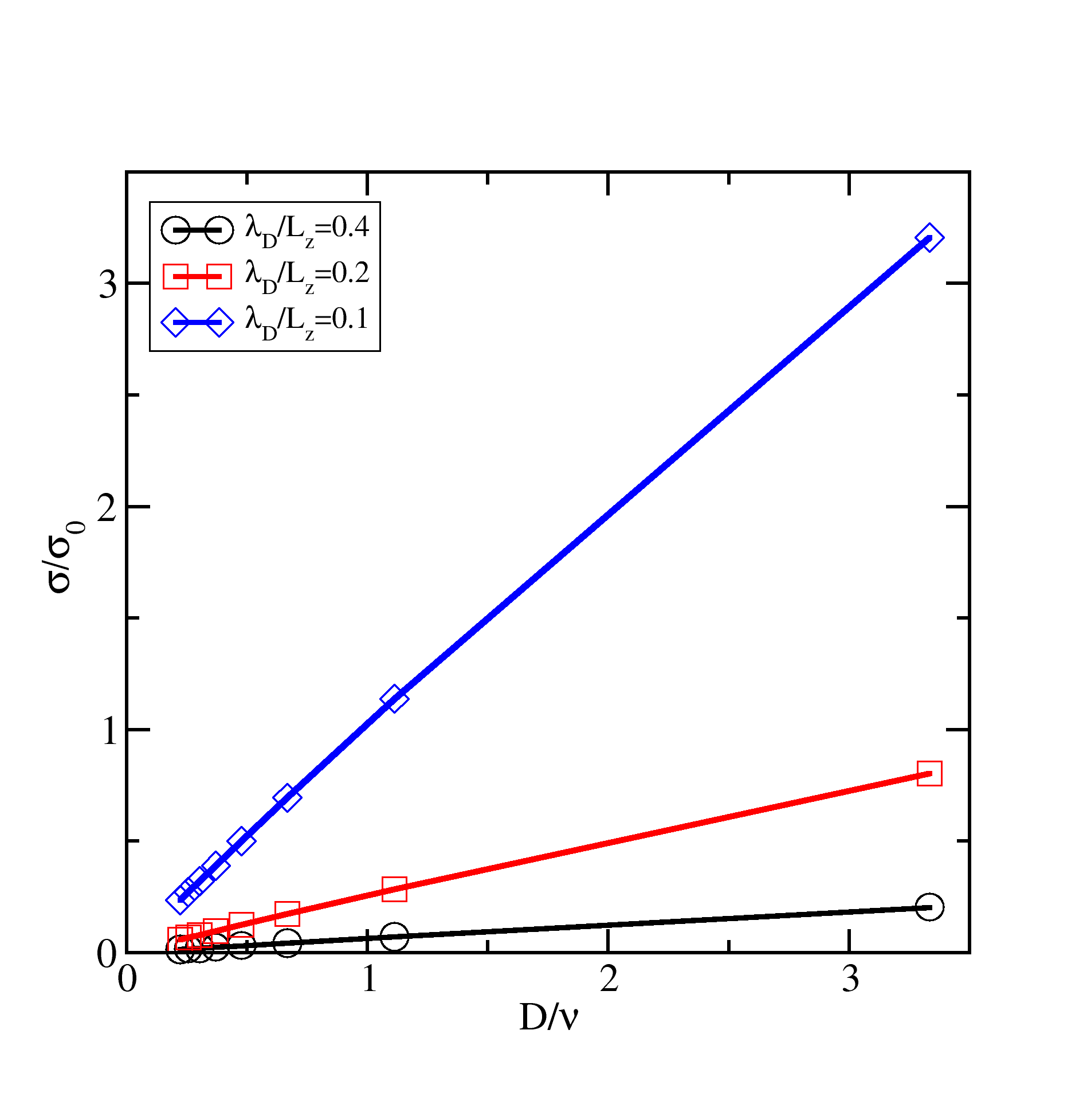}
\caption{
Electric conductivity for the charged slit channel as obtained via the LBM
for $\lambda_D/L_z=0.4$ (circles), $0.2$ (squares) and $0.1$ (diamonds).
The linear dependence of the electric current on diffusivity
reflects the dominance of the Ohmic contribution,
which is about $100$ times larger than the convective counterpart.}
\label{condvsigma}
\end{center}
\end{figure}

\begin{figure}[htbp]
\begin{center}
\includegraphics[scale=.8]{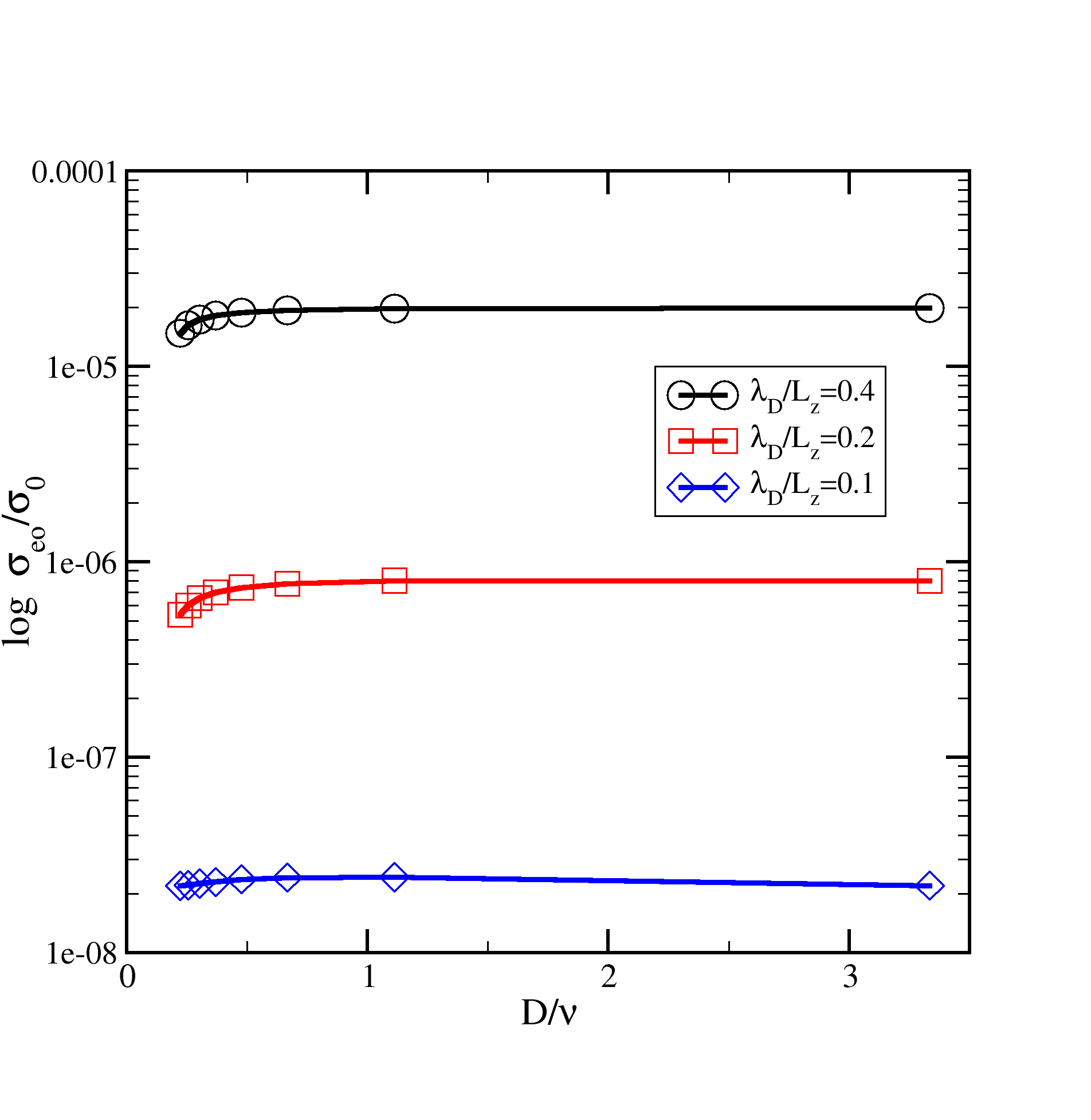}
\caption{Electro-osmotic contribution to the electric conductivity (see Eq.~\eqref{curreo}
and text for details) for the charged slit channel
for $\lambda_D/L_z=0.4$ (circles), $0.2$ (squares) and $0.1$ (diamonds).
}
\label{eovsdiff}
\end{center}
\end{figure}

The behavior of the  curves in Fig.~\ref{condvsigma} is understood by noting 
that the number of charge carriers in the mixture is inversely proportional 
to $\lambda_D^2$ \cite{masliyah2006electrokinetic}. 
Again, the electro-osmotic charge flow turns out to be an increasing function of $\lambda_D$
as shown in Fig.~\ref{eovsdiff},
but depends on the mutual diffusivity, in particular by displaying a slight increase with $D$.
The reason for such behavior is the same as the one mentioned above. 

Having different velocities between the solvent and the barycentric one 
at small $\omega_{drag}$, that is at high diffusivity, 
gives rise to a larger barycentric velocity. We performed a
theoretical analysis of this occurrence at the level of linearized hydrodynamics
starting from the kinetic equation \eqref{evolution}. The results show
a fairly good agreement with the simulations data and
predict that the phenomenon is only visible at small values of $\omega_{drag}$.
For the sake of conciseness, we will report the calculations elsewhere.

In essence, the Lattice Boltzmann  algorithm,  derived from the kinetic model introduced 
in this work, is capable  not only to accurately reproduce the behavior
predicted by  the continuum equations of Sec.~\ref{macro},
but also to exhibit non-trivial features of electrokinetic transport
arising from the interplay of diffusion, convection and electrostatics.

\section{Conclusions}
\label{Conclusions}

To summarize, we modified the BGK dynamics in order to control separately 
the timescales associated to concentration and momentum diffusion. 
The presented method is derived from microscopic considerations involving the exact treatment 
of the collision term in the Boltzmann equation. We replaced the microscopic expression 
for the inter-species frictional force, which tends to equalize the species velocities, 
by a simpler effective term depending on a tunable frequency $\omega_{drag}$. 
The derived equations have similarities with the ones put forward by Luo and Girimaji
\cite{luo2003theory}. However, the orthogonalization procedure described here
represents a key ingredient to separately control concentration and momentum diffusion. 

We further implemented the equations in the context of the Lattice Boltzmann Method, 
and numerically verified  that controlling independently the diffusivity 
and the kinematic viscosity for different physical models is feasible and
mirrors the expected dynamics in the continuum.
We applied the method to the numerical calculation of the diffusion coefficient 
in neutral binary mixtures and of the electric currents of ternary charged mixtures 
under uniform  bulk conditions, finding a perfect agreement with the theoretical prediction.

Under non uniform conditions, such as those realized in a charged slit geometry,
the electro-osmotic current exhibits a non-trivial behavior, since
the convective contributions to the mass and charge currents display 
an interesting dependence on the diffusion coefficient. 
Such occurrence, to the best of our knowledge, was not previously reported in the literature.

We conclude with a remark concerning the relation between the present approach 
and the standard single relaxation, BGK description of mixtures. 
When the value of the tunable parameter $\omega_{drag}$ is chosen
to be equal to $\omega_{visc}$, on physical grounds one does not expect to observe 
differences between the two methods. Not only we have confirmed numerically
such an occurrence, but it we also demonstrated that the kinetic equation 
\eqref{evolution} in practice reduces to the standard BGK single relaxation form 
in this limit.

\subsection{Acknowledgments}
This work was supported by the Italian Ministry of University and Research 
through the ``Futuro in Ricerca'' project RBFR12OO1G - NEMATIC.

\clearpage

\appendix
\section{Hermite expansion of the kinetic equation \eqref{evolution}}

The standard procedure to construct the LBM algorithm starting from the kinetic 
equation is to expand each term of the equation in the Hermite polynomial basis. 
It is apparent that the coefficients of the expansion of the distribution function
and the collisional terms are combination of the macroscopic fields \cite{shan2006kinetic}. 
This property, together with the orthogonality of the basis set, 
not only guarantees that, given an expansion order $K$, the first $K$ moments 
of the distribution are untouched by the truncation but also permits to control 
the error introduced by the evolution equation in the macroscopic 
fields \cite{martys1998evaluation,shan2006kinetic}. 
Using the Einstein convention on the repeated indices, the one particle distribution function 
expands as
\bea
&&f(\rr,\vv,t)=\Gamma_0(\vv)\sum_{k=0}^{\infty}{1\over v_T^{2k}k! }a^k_\ii(\rr,t){\cal H}^{k}_{\ii}(\bv)\simeq \Gamma_0(\vv)\sum_{k=0}^{K}{1\over v_T^{2k} k!}a^k_\ii(\rr,t)   {\cal H}^{k}_{\ii}(\bv)\equiv f^K(\rr,\vv,t)  \;;\;\;\nonumber\\
&&\ii=\{i_1,..,i_k\}\;;\;\;i_l=\{1,2,3\}\;\;\;\;\Gamma_0(\bv)={1\over {(2\pi v_T^2)}^{3/2}}\exp({-{\bv^2\over 2v_T^2}})\;\;\;;\;\;\;\;\;\; {\cal H}^{k}_{\ii}(\bv)={(-1)^k\over \Gamma_0(\bv)}\NN^k_{\ii}\Gamma_0(\bv)\;\; ; 
\eea
where $\NN^k_{\ii}$ represents the k-th derivative with respect to $v_{i_1},..,v_{i_k}$ and
$\delta^n_{\ii\jj}$ is a sum of products of $n$ $\delta$'s, each $\delta$ having one 
subscript from the set $\ii$ and one from the set $\jj$ 
\cite{grad1949note}. 
From the definition, it follows that
\bea
a^k_\ii(\rr,t) = 
 \int f(\rr,\bv,t){\cal H}^{k}_{\ii}(\bv)d\vv 
\equiv <f|{\cal H}^{k}_{\ii}> 
\label{norherm}
\eea
In the following, we will often make use of the identities 
${\cal H}^2_{ij}(\vv)={\cal H}^2_{ij}(\vv-\uu)+\uu_i(\uu-\vv)_j+\uu_j(\uu-\vv)_i+\uu_i\uu_j$
and consider the  Hermite expansion  up to second order of the equation
\be
  \frac{\partial}{\partial t}\fa(\rr,\vv,t) +\vv\cdot\NN \fa(\rr,\vv,t) 
 = -\omega_{visc}[ \fa(\rr,\vv,t)- \phi^{\alpha}_{\perp}(\rr,\vv,t)]+g^{\alpha}(\rr,\vv,t)
 \label{galfa}
\ee
with 
$$
g^{\alpha}(\rr,\vv,t)\equiv \frac{{\bf F}^{\alpha,drag}(\rr,t) }{m v_{T}^2} \cdot(\vv-\uu(\rr,t)) \phi_{\alpha}(\rr,\vv,t) .
$$
Both terms  in the r.h.s. of Eq.~\eqref{galfa} have a dependence of the type 
$(\vv-\uu)\phi^{\alpha}$, so that
\bea
&&\bra g^{\alpha}|{\cal H}^{0}_0\ket={1\over v_T^2}\int{{\bf F}^{\alpha,drag}}\cdot (\bv-\uu)\phi^{\alpha}d\bv =0
\nonumber\\
&&\bra g^{\alpha}|{\cal H}^1_i\ket={1\over v_T^2}\int\bv_i \Big({{\bf F}^{\alpha,drag}} \cdot(\vv-\uu)\Big)\phi^{\alpha}d\bv=
{1\over v_T^2}\sum_j {{\bf F}_j^{\alpha,drag}}S^1_{ij}
\nonumber\\
&& \bra g^{\alpha}|{\cal H}^2_{ij}\ket= {1\over v_T^2}\sum_k{{\bf F}_k^{\alpha,drag}}\int (\bv_i\bv_j-v_T^2\delta_{ij})(\vv-\uu)_k\phi^{\alpha}d\bv\equiv{1\over v_T^2}\sum_k{{\bf F}_k^{\alpha,drag}}S^1_{ijk}
\label{firstintegrals}
\eea
and
\bea
&&S^1_{ij} =\int(\bv -\uu)_i(\vv-\uu)_j\phi^{\alpha}d\bv+{\uu}_i\int(\vv-\uu)_j\phi^{\alpha}d\bv=v_T^2\delta_{ij}
\nonumber\\
&& S^1_{ijk}=\int(\bv_i\bv_j-v_T^2\delta_{ij}) (\bv-\uu)_k\phi^{\alpha}d\bv=v_T^2(\uu_i\delta_{jk}+\uu_j\delta_{ki})
\eea
The force term has the following discrete representation
\bea
\frac{{\bf F}^{\alpha,drag}(\rr,t) }{v_{T}^2} \cdot(\cc_p-\uu(\rr,t)) 
\phi_{\alpha}(\rr,\cc_p,t)
&\simeq&
n^{\alpha}(\rr)\Gamma_0(\cc_p)\Big[{1\over v_T^4}\sum_{ij} {{\bf F}_j^{\alpha,drag}}S^1_{ij}{\cal H}_i^1(\cc_p)+{1\over 2v_T^6}\sum_{ijk}{{\bf F}_k^{\alpha,drag}}S^1_{ijk}{\cal H}^2_{ij}(\cc_p)\Big]
\nonumber\\
&=&n^{\alpha}(\rr)\Gamma_{0p}\Big[{\FF^{drag}\cdot \cc_p\over v_T^2}+{1\over  v_T^4}\Big( (\FF^{\alpha,drag}\cdot\cc_p)(\uu\cdot\cc_p)-v_T^2\FF^{\alpha,drag}(\rr,t)\cdot\uu\Big) \Big]
\nonumber\\
&\equiv& {{\bf g}_p^{\alpha}(\rr,t)}\nonumber\\
\label{ourforc}
\eea
where $\Gamma_{0p}=\Gamma_{0}(\cc_p)$.
The analogous expansion of $\phi^{\alpha}_{\perp}$ reads
\bea
\phi^{\alpha}_{\perp}(\rr,\vv,t)
&=&
\phi^{\alpha}(\rr,\vv,t) \Bigl\{1+
{{1\over{v_{T}^2}}(\uua-\uu)\cdot {\cal H}^1(\vv-\uu)}+{1\over 2v_T^4}(\uua-\uu)(\uua-\uu):{\cal H}^2(\vv-\uu)\Bigl\}\nonumber\\
&=&\phi^{\alpha}(\rr,\vv,t)+{1\over{v_{T}^2}}(\uua-\uu):\phi^{\alpha}(\rr,\vv,t){\cal H}^1(\vv-\uu)+{1\over{2v_{T}^4}}(\uua-\uu):\phi^{\alpha}(\rr,\vv,t){\cal H}^2(\vv-\uu)
\eea

The expansion of $\phi^{\alpha}$ to second order is well-known 
(see for example \cite{shan2006kinetic}), so that we report here only the projections of 
$\phi^{\alpha}{\cal H}^1(\vv-\uu)\,$,  $\phi^{\alpha}{\cal H}^2(\vv-\uu)$ to
the same order. The integrals involved in the calculation 
of the projection {of $\phi^{\alpha}{\cal H}^1(\vv-\uu)\,$,} were given before (see Eq.~\eqref{firstintegrals}) 
while for the remaining ones, we obtain:
\bea
&&S^2_{ij}=\int \phi^{\alpha}[(\vv-\uu)_i(\vv-\uu)_j-v_T^2\delta_{ij}]d\vv=0\nonumber\\
&&S^2_{ijk}=\int \vv_k\phi^{\alpha}[(\vv-\uu)_i(\vv-\uu)_j-v_T^2\delta_{ij}]d\vv=0\nonumber\\
&&S^2_{ijkl}=\int (\vv_l\vv_k-v_T^2\delta_{lk})\phi^{\alpha}{\cal H}^2_{ij}(\vv-\uu)d\vv=
v_T^4(\delta_{il}\delta_{jk}+\delta_{ik}\delta_{jl})
\eea
Putting all together, 
\bea
\phi^{\alpha}_{p,\perp} = 
n^{\alpha}\Gamma_{0p}
\Big\{
1+\frac{\uu^{\alpha}\cc_p}{v_T^2}+\frac{1}{2v_T^4}
\Big[2(\uu^{\alpha}\cdot\cc_p)(\uu\cdot\cc_p)-2v_T^2(\uu^{\alpha}\cdot\uu)-(\uu\cdot\cc_p)^2+v_T^2|\uu|^2
+\Big((\uua-\uu)\cdot\cc_p\Big)^2-v_T^2|\uua-\uu|^2
\Big]
\Big\}
\nonumber \\
\eea

\clearpage

\bibliographystyle{unsrt}

\end{document}